# Investigation of Rule 73 as Case Study of Class 4 Long-Distance Cellular Automata

Lucas Kang 2013

**Abstract**  ────────────────────────────────────────────────────

Cellular automata (CA) have been utilized for decades as discrete models of many physical, mathematical, chemical, biological, and computing systems. The most widely known form of CA, the elementary cellular automaton (ECA), has been studied in particular due to its simple form and versatility. However, these dynamic computation systems possess evolutionary rules dependent on a neighborhood of adjacent cells, which limits their sampling radius and the environments that they can be used in.

The purpose of this study was to explore the complex nature of one-dimensional CA in configurations other than that of the standard ECA. Namely, "long-distance cellular automata" (LDCA), a construct that had been described in the past, but never studied. I experimented with a class of LDCA that used spaced sample cells unlike ECA, and were described by the notation LDCA-x-y-n, where x and y represented the amount of spacing between the cell and its left and right neighbors, and n denoted the length of the initial tape for tapes of finite size. Some basic characteristics of ECA are explored in this paper, such as seemingly universal behavior, the prevalence of complexity with varying neighborhoods, and qualitative behavior as a function of x and y spacing.

Focusing mainly on purely Class 4 behavior in LDCA-1-2, I found that Rule 73 could potentially be Turing universal through the emulation of a cyclic tag system, and revealed a connection between the mathematics of binary trees and Eulerian numbers that might provide insight into unsolved problems in both fields.

**Introduction** ─────────────────────────────────────────────────────────────

Cellular automata (CA) have been used for decades as mathematical idealizations of physical systems, in which space and time are discrete, and in studies of "self-organizing" physical, chemical, and biological phenomena [1][2][3][4][5][6]. In the world of computational science, several classifications of CA have risen to prominence, ranging from one to three dimensional automata to those of even higher dimension. (A survey of CA classifications is available in [24].) However, Wolfram's classification of elementary cellular automata (ECA) have become one of the most widely known forms of evolving computational systems, and have arguably revolutionized the exploration of cellular automata [2]. The purpose of this paper is to study the complexity of one-dimensional cellular automata in configurations other than that of Wolfram's elementary cellular automata. I explore a new class of "long-distance cellular automata" (LDCA) whose properties are not governed by ECA, and demonstrate that certain LDCA possess interesting qualities and are candidates for computational universality, which implies that they may be used to solve problems in other fields of research [2][7][8].

ECA are defined as the simplest class of one-dimensional cellular automata, and have two possible states for each cell (0 or 1). These states are placed in an "array", so that every "cell" dictates the condition of a particular part of the system. This array, which can be finite or infinite in length, when taken from a one-dimensional CA or ECA can be thought of as resembling a tape of a Turing machine, so it will be referred to as such through the extent of this paper [2][3]. The evolution of an array is the collective evolution of all individual cells, which is dependent only on the values of its "neighborhood". The neighborhood is defined as containing the cell itself and all immediately adjacent sites; in one-dimensional cellular automata, the neighborhood consists of 3 adjacent cells. As a result, the evolution of any elementary cellular automaton can be described through a table of evolutionary steps, specifying the state a given cell will have in the next generation based on the values of the cell to its left, the cell to its right, and the target cell itself. Since for every ECA, there must exist $2^3$=8 such steps in a table, and since for every step there are 2 possible outcomes (0 or 1), there exist a total of $2^8$=256 total tables to describe ECA. Each of these tables is said to be a "Rule" of evolution and so there exist 256 defined ECA, ranging from Rule 1 to Rule 256 [2].

Long-distance cellular automata (LDCA), an extension of ECA rules, use spaced sample cells unlike ECA, and are described by notation LDCA-x-y-n, where x and y represent the amount of spacing between the cell and its left and right neighbors, and n denotes the length of the initial tape (for tapes of finite size). The values for x and y must always be greater than 0, and since the tape of a finite LDCA is cyclic, x + y is unrestricted. In this experiment, I used LDCA-1-2, so as to form a basis for a systematic exploration with larger x and y values.

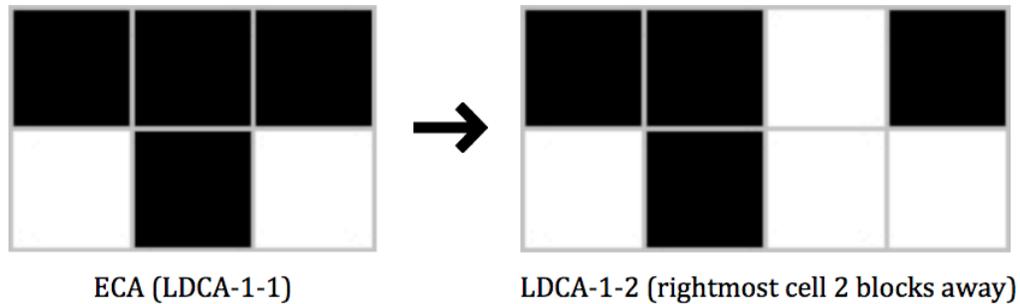

Figure 1: The layout of the three sampled cells and one output cell in an ECA (left), and a corresponding image for the configuration of LDCA-1-2-n (right).

LDCA, due to their separated sample cells, provide functions that ECA cannot, and theoretically can be used to model chaotic systems in which every evolutionary step possesses an inherent random input. This is the case because as an ECA rule is applied in configurations LDCA-x-1 or LDCA-1-y on a tape of random initial conditions, for 1 < x or y < n-4, it behaves as if it were a two-cell CA, while the third, more remote cell provides a seemingly unrelated and potentially random source of input. We can see this form of evolution, in which an outside effect or random input actively modifies a system, in maps of electrical activity where negligible but existent charges affect components, biological systems such as angiogenesis where capillary growth is predictable but non-deterministic, and chaos theory's butterfly effect in which a relatively small initial change can affect a larger outcome. Additionally, when the cellular automaton's configuration approaches LDCA-x-y, for $1 < x$ or $y < n-4$, the result can be unrecognizable in comparison to the standard ECA evolution of the same rule.

LDCA also have the distinct property of possessing cyclic state transition diagrams. When applying an ordered list of all 256 ECA rules to an LDCA-x-y-n such that $x + y = 0$ (mod n) or $x + y = x$ or $y$ (mod n), then the rules will generate a series of repeated state transition diagrams. If the LDCA samples only the target cell and one other cell, so that $x + y = 0$ (mod n), then the table of rules will create two distinct list of diagrams with 4 rules each that will be alternated. If the LDCA samples one cell from anywhere on the tape, and one from the same cell as the center cell, so that $x + y = x$ or $y$ (mod n), then there are two possible outcomes. If $x = n$ and therefore $x + y = y$ (mod n), then the repeated lists of state transition diagrams will be 4 rules in length. If $y = n$ and so $x + y = x$ (mod n), then the lists will be 8 rules in length. And if the LDCA samples all three cells from the same cell so that $x + y = 0$ (mod n) and $-x + y = 0$ (mod n), then each repeated list of diagrams will have 2 rules each.

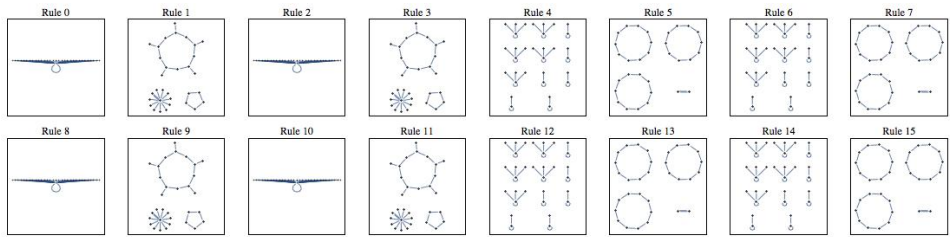

Figure 2: ECA rules in LDCA-1-4-5. Since $x + y = 0$ (mod n), the state transition diagrams repeat in groups of 4 [11].

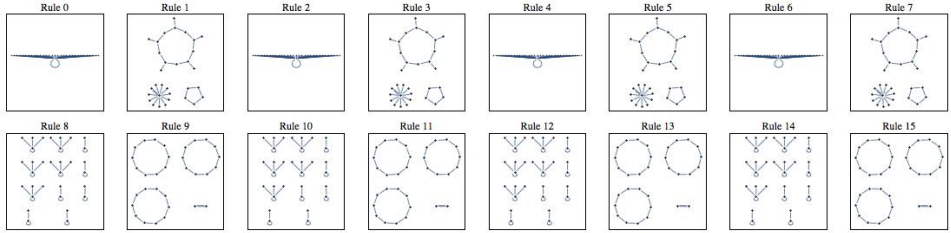

Figure 3: ECA rules in LDCA-1-5-5. Since $x + y = x \pmod{n}$, the state transition diagrams repeat in groups of 8 [11].

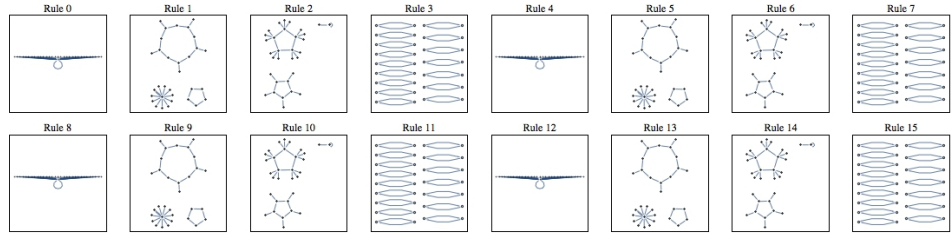

Figure 4: ECA rules in LDCA-5-1-5. Since $x + y = y \pmod{n}$, the state transition diagrams repeat in groups of 4 [11].

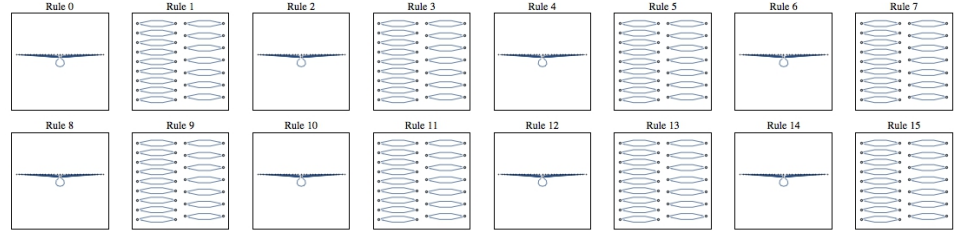

Figure 5: ECA rules in LDCA-5-5-5. Since $x + y = 0 \pmod{n}$ and $-x + y = 0 \pmod{n}$, the state transition diagrams repeat in groups of 2 [11].

## Rules 73 and 109

The 2-color, 2-state cellular automaton we will study in the LDCA-1-2 configuration is known as "Rule 73" according to Wolfram's numbering scheme [2]. In the evolution of Rule 73, each cell is in one of two states $\{0, 1\}$, and since the rule is being applied to a LDCA-1-2 configuration, at each discrete time step every cell synchronously updates itself according to the value of itself and its nearest neighbors: $F(C_{i-1}, C_i, C_{i+2})$, where F is the following function [12]:

$F(1, 1, 1) = 0$    $F(1, 1, 0) = 1$    $F(1, 0, 1) = 0$    $F(1, 0, 0) = 0$

$F(0, 1, 1) = 1$    $F(0, 1, 0) = 0$    $F(0, 0, 1) = 0$    $F(0, 0, 0) = 1$

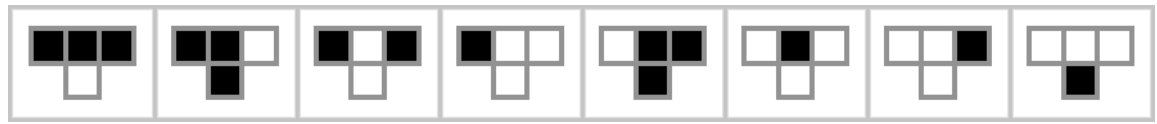

Figure 6: This table depicts the evolutionary substitution "rules" of Rule 73. Rule 73 and Rule 109 are equivalent through both left-right and color equivalence.

Interestingly, Rule 73 is equivalent to Rule 109, through both left-right and color equivalence, which means that the two rules are identical after either one's color is inverted and evolution is mirrored horizontally. This entails that studying either rule implies the exploration of the other. Rule 109 evolves according to the function $G(C_{i-1}, C_i, C_{i+2})$, where G is the following function [13]:

$G(1, 1, 1) = 0$  $G(1, 1, 0) = 1$  $G(1, 0, 1) = 1$  $G(1, 0, 0) = 0$

$G(0, 1, 1) = 1$  $G(0, 1, 0) = 1$  $G(0, 0, 1) = 0$  $G(0, 0, 0) = 1$

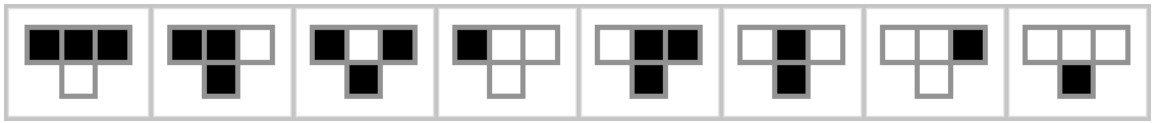

Figure 7: This table depicts the evolutionary substitution "rules" of Rule 109. Rule 109 and Rule 73 are equivalent through both left-right and color equivalence.

*Computational Universality* ─────────────────────────────────────────

Universality was a major factor in the choosing of Rule 73; it was suggested that "Class 4" cellular automata might be capable of universal computation in 1984 [8], and Rule 73 is one of two purely-Class 4 automata in LDCA-1-2, with the other being Rule 109 and therefore equivalent. (Both Rule 73 and Rule 109 are defined as Class 2 Rules in the ECA rule space, but when evolved as LDCA-1-2, they exhibit purely Class 4 behavior. They have also been shown to possess Class 4 behavior in other contexts [25].) Besides exploring the properties and characteristics of Rule 73, this study will also attempt to demonstrate the universality of Rule 73.

Universal computational systems are those that are theoretically capable of emulating any other system [2][8][14]. This means that a singular system would be capable of behaving as any other mathematically definable system, which has significant implications in computational science. Such systems usually require an encoding and decoding process, in order to translate information and behavior [14]. For example, Boolean logic systems, or computer programs, are universal, but only after the system being emulated has been coded in binary, and the result of the program translated back into the language of the original system.

The proving of a computational system's universality is usually done through the emulation of another system, previously known to be universal. As a result of the Church-Turing Thesis [15], Turing machines have been defined as universal. Then, in 2004, Cook proved that cyclic tag systems could successfully emulate universal Turing machines, and were therefore universal [2][14]. While several cellular automata, have been shown or suspected to be universal, the most commonly known example is that of the elementary cellular automata Rule 110, which was shown to emulate a universal cyclic tag system [14].

*Methodology: Visualization* ────────────────────────────────

The shift from ECA to LDCA-1-2 brings several changes to the Rule 73, including the way that it is visualized. When the evolution of Rule 73 in LDCA-1-2 is plotted normally, gliders, or "particles" as they will be referred to in this paper, tend to blend into the background, or become difficult to distinguish as they collide.

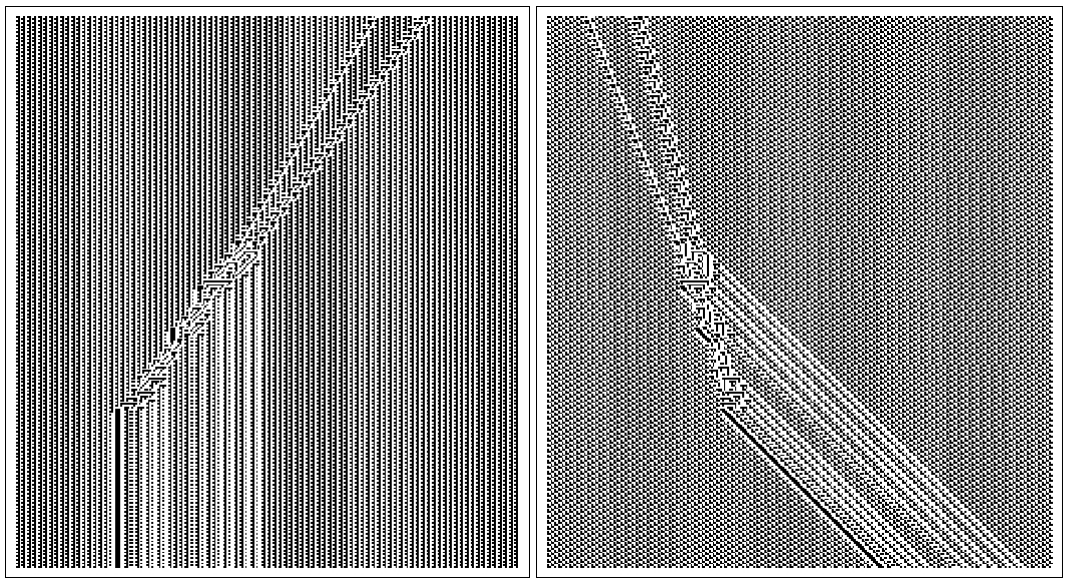

Figure 8: Here we can see that LDCA 73 possesses several stationary particles (vertical lines in picture on the left). When the plot of Rule 73's evolution is skewed however, collisions between gliders and the resulting particles become more apparent.

While it is standard to plot the tape of a cellular automaton so that cell *i* of a previous state is directly above cell *i* in the next state, we will skew the evolution of LDCA-1-2 so that the position of cell *i* in time step *x* corresponds to that of cell *i-1* in *x+1*. By applying a rotation function to the CA, we can more clearly utilize the methods described in this paper:

*Table*[*Nest*[*RotateRight*, #[[i]], i - 1], {i, 1, Length[#]}] & *CellularAutomaton*[…]

In this study, we will assume that all calculations and plots regarding Rule 73 in LDCA-1-2 are developed with an *i* → *i-1* skew.

*Methodology: Block Emulation* ─────────────────────────────────

Block emulation is a form of emulation that can be used to find emulations of cellular automata in different rule spaces. By substituting certain blocks in a certain CA, one can create a simpler or more complicated automaton, which may emulate other rules, or a different computational system. The main idea is to encode one cell of Rule A into n blocks of cells of Rule B. By replacing corresponding blocks according to a set of rules, one can transform a cellular automaton of one rule into a replica of another rule [2]. With this concept, one can show for example that Rule 22 is able to emulate Rule 90. However, this type of emulation is not possible in all cases. Some rules with blocks up to a certain block size, Rule 30 for example, are not able to emulate any fundamental rules at all through block emulation.

In an effort to see what Rule 73 in LDCA-1-2 emulates, we can use block emulation in the 3/2 rule space. To begin, we must convert Rule 73 to a rule in the 3/2 rule space, so we find all evolutionary rules in Rule 73 that return a black cell:

$F(1, 1, 1) = 0$     $F(1, 1, 0) = 1$     $F(1, 0, 1) = 0$     $F(1, 0, 0) = 0$

$F(0, 1, 1) = 1$     $F(0, 1, 0) = 0$     $F(0, 0, 1) = 0$     $F(0, 0, 0) = 1$

Then, find all corresponding configurations of cells in the 3/2 rule space. In 3/2, the cells are formatted as *a, b, c, d*, with the cellular automata sampling cells *a, b,* and *d*. Since cell *c* remains unsampled, it can be either 1 or 0 in our translation from ECA to 3/2. Therefore, we end up with:

$$F_{3/2}(1, 1, 0, 0) = 1 \qquad F_{3/2}(1, 1, 1, 0) = 1$$

$$F_{3/2}(0, 1, 0, 1) = 1 \qquad F_{3/2}(0, 1, 1, 1) = 1$$

$$F_{3/2}(0, 0, 0, 0) = 1 \qquad F_{3/2}(0, 0, 1, 0) = 1$$

Next, we must convert all of the possible inputs into base 10:

$$(1, 1, 0, 0) \rightarrow 12 \qquad (1, 1, 1, 0) \rightarrow 14$$

$$(0, 1, 0, 1) \rightarrow 5 \qquad (0, 1, 1, 1) \rightarrow 7$$

$$(0, 0, 0, 0) \rightarrow 0 \qquad (0, 0, 1, 0) \rightarrow 2$$

And sum 2 to the power of each of the results to get the rule number of 20645:

$$2^{12} + 2^{14} + 2^5 + 2^7 + 2^0 + 2^2 = 20645$$

Finally, a comparison between Rule 20645 in 3/2 and Rule 73 in ECA confirms that they are identical rules:

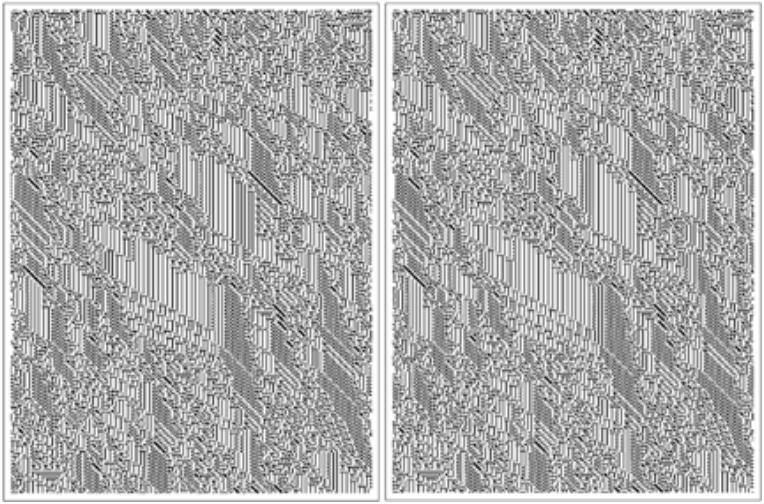

Figure 9: A comparison between Rule 73 in LDCA-1-2 (left) and Rule 20645 in the 3/2 rule space (right) with equivalent initial conditions. These two rules are identical (hence the identical evolutions above) and can be derived from one another by converting from an ECA rule space to the 3/2 rule space and vice versa. As always, the LDCA-1-2 is plotted with an
*i → i-1* skew.

*Methodology: Neighbor-Dependent Substitution System* ⎯⎯⎯⎯⎯⎯⎯⎯⎯⎯⎯⎯⎯⎯⎯⎯

Substitution systems form the backbone of most computational systems, but cannot in general emulate cellular automata. However, when substitution systems have rules that depend not only on the color of a single element, but also on the color of at least one of its neighbors, they display more complicated behavior [2]. In order to utilize this behavior, one must be able to identify the current state of a CA with an array or string of values, and construct a substitution system for its behavior that is closed under evolution (so that as substitutions occur, no results arise that cannot be interpreted). "Neighbor-dependent substitution systems" are known to emulate certain cellular automata, and could be helpful in proving universality of Rule 73 [2].

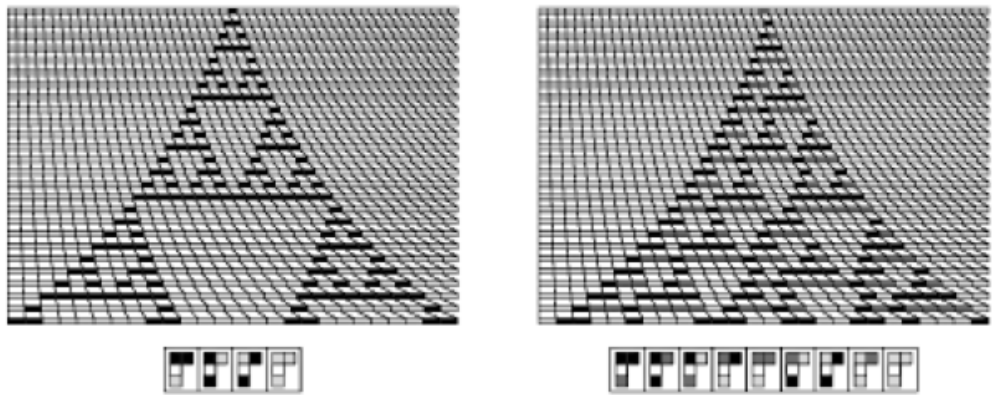

Figure 10: Neighbor-dependent substitution systems that emulate Rule 90 (left) and 30 (right). The systems shown are examples of neighbor-dependent substitution systems with highly uniform rules that always yield one cell and correspond directly to known cellular automata [11].

*Results: Characteristics of Rule 73* ⎯⎯⎯⎯⎯⎯⎯⎯⎯⎯⎯⎯⎯⎯⎯⎯⎯⎯⎯⎯⎯⎯

After a brief exploration of Rule 73, I found several interesting phenomena, two of which are below. Firstly, for Rule 73 being evolved in an LDCA-1-y configuration, as y increases in value from 1, complex behavior arises at y = 2 and is extinguished after y = 3, giving rise to pseudo-chaotic behavior. An analysis of Rule 73's behavior for y > 3, revealed that the Gaussian distribution of all possible states of Rule 73 was not normally distributed. This implies that the chaotic nature of Rule 73 at high values of y is not random, but rather, is complex.

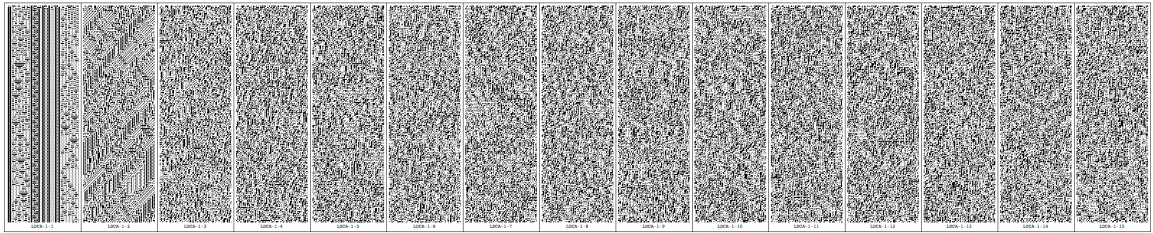

Figure 11: While evolving Rule 73 in a LDCA-1-y configuration, as y increases in value from 1, complex behavior arises and is extinguished, giving rise to chaotic behavior that resembles white noise. The Gaussian distribution of all possible states is not normally distributed in these chaotic states, which implies that the chaotic nature of Rule 73 at high values of y is not random, but is complex and difficult to perceive.

I also found another effect of increasing the value of y. When viewing the state transition diagrams, one notices a repetition of structure, not unlike that mentioned in the introduction.

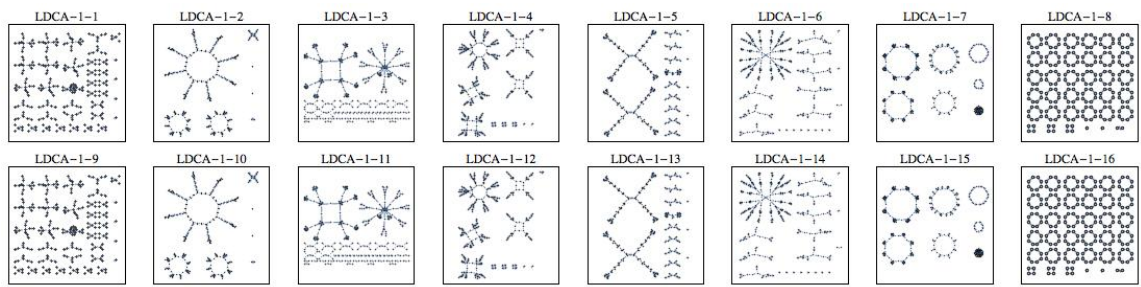

Figure 12: As Rule 73 is applied in different configurations the effect of having a high y value becomes apparent through state transition diagrams. Given Rule 73 applied as LDCA-1-y-n, the state transition diagrams are repeated with a period of n as the y value increases since the tape is cyclic [11].

**Characteristics of Rule 73: Particles**

The gliders or "particles" that exist in Rule 73 move with 4 different velocities (0, 1/4, 2/5, and 1) over a constant background. A single "block" of background is represented by "101100", "110010" or "001011", has an evolutionary period of 3, and spatial period of 6. In the naming convention for Rule 73 in LDCA-1-2, a single block of background is represented with a " - ", half of a background block is denoted with a " ' ", and a single particle is a letter (its name).

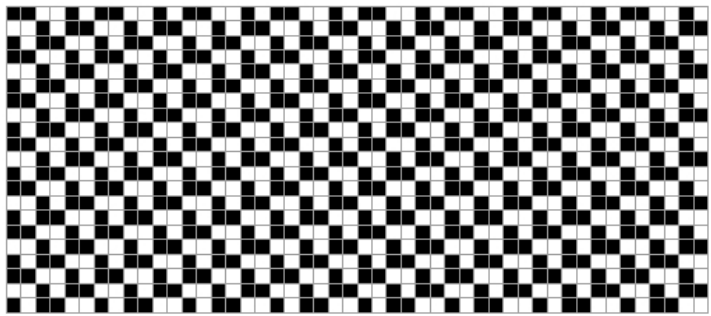

Figure 13: An image of the background of Rule 73 in LDCA-1-2. The background resembles a surface tessellated with "L" shaped units; a single row of background is represented by "101100", "110010" or "001011", and has period 3 (vertically) and spatial period 6 (horizontally). The above image is scaled and partitioned, so that one can better see the structures of cells in the background.

There are several particles in Rule 73 that act as the building blocks for larger constructs and "compound particles." These are called fundamental particles, and are the main focus of this exploration. All fundamental particles are organized and labeled by velocity and phase shift (or mass), with the mass ranging in value from 0 to +6, and representing the number of cells that the background is shifted to the right by the presence of the particle.

| Particle | Velocity | Mass | Period | Particle | Velocity | Mass | Period |
|---|---|---|---|---|---|---|---|
| A | 0 | 0 | 3 | F | 1 | 2 | 2 |
| B | 0 | 5 | 3 | Fbar | 1 | 3 | 4 |
| C | 1/4 | 1 | 8 | G | 1 | 0 | 2 |
| D | 2/5 | 4 | 5 | H | 1 | 3 | 2 |
| E | 2/5 | 2 | 15 | | | | |

Figure 14: Table of all fundamental particles in Rule 73 in the LDCA-1-2 configuration. All fundamental particles are organized and labeled by velocity, phase shift (or mass), and period. The mass ranges in value from 0 to +6, and represents the number of cells that the background is shifted to the right by the presence of the particle.

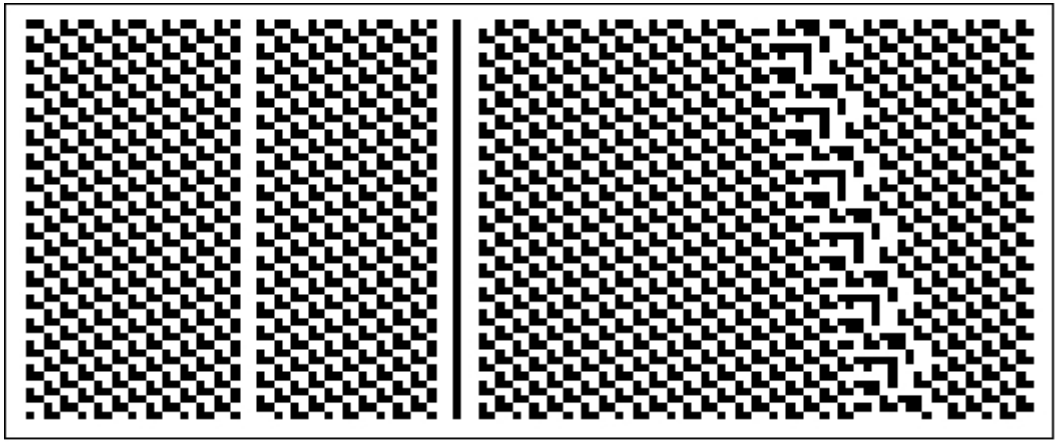

Figure 15: Particles A, B, and C; velocities 0, 0, and 1/4.

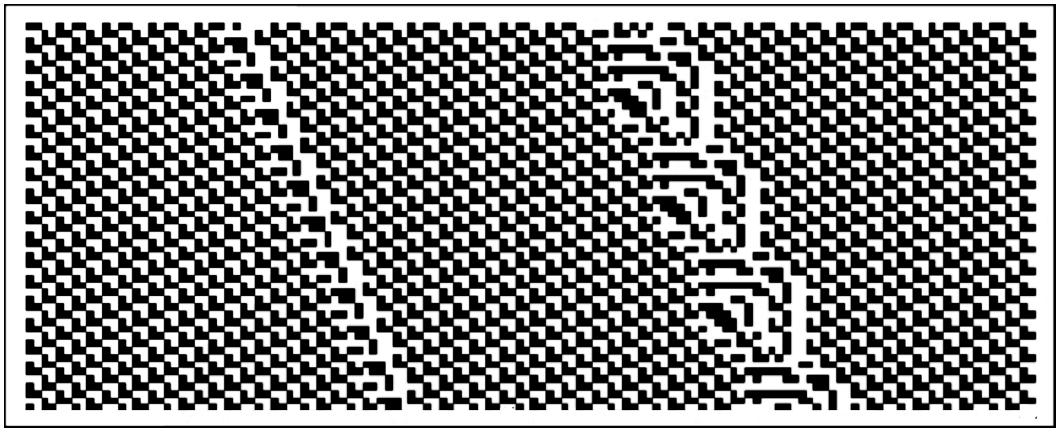

Figure 16: Particles D and E; velocities 2/5, 2/5.

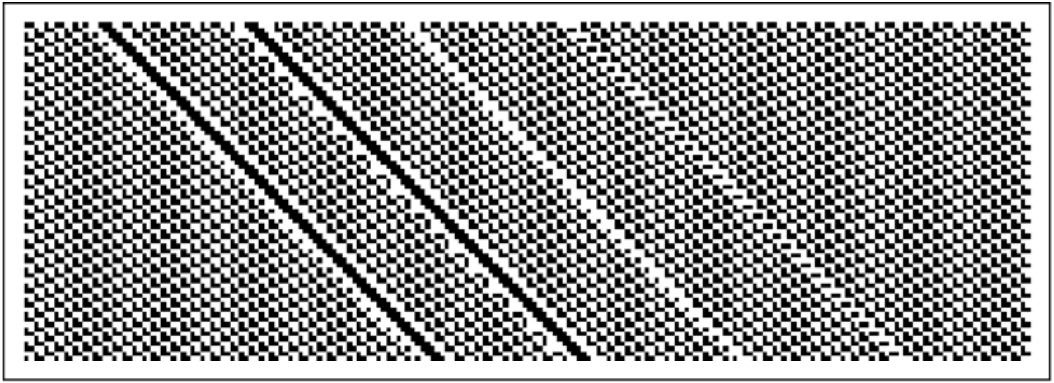
Figure 17: Particles F, Fbar, G, H; velocities of 1 for all.

However, merely placing 2 fundamental particles next to each other on a tape cannot create certain compound particles. For example, in the cases of G'G and G'H, the second particle in the pair must be shifted vertically by one evolutionary step.

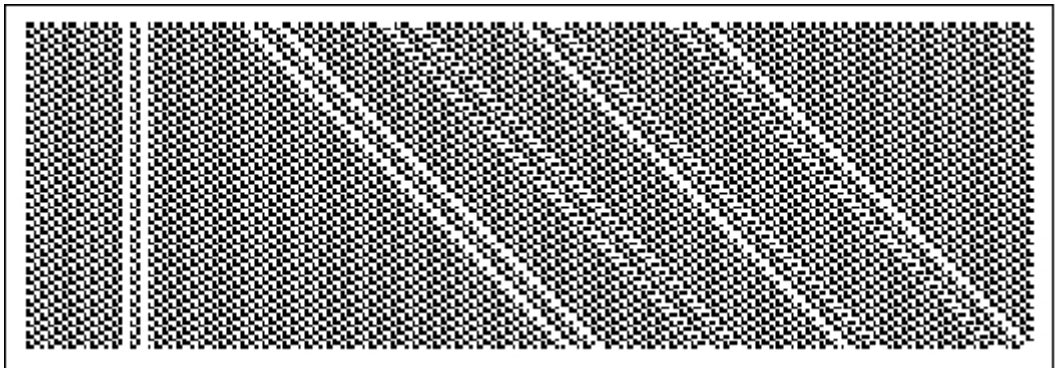
Figure 18: Particles B'B, G'G, H'H, G'H, H'G. These are unique compound particles that must be formed by vertically shifting one of the two particles before placing them adjacent.

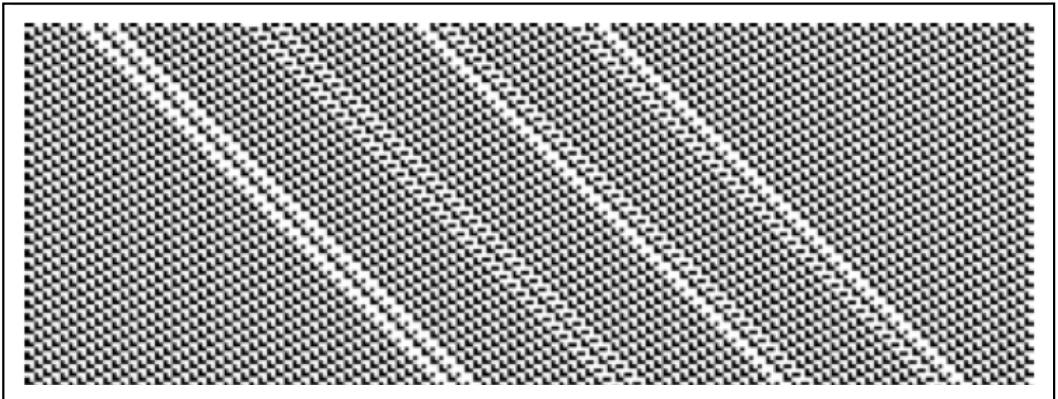
Figure 19: Particles GG, HH, GH, HG. These are documented due to their prevalence in collisions and systems made from Rule 73.

## *Results: Collisions* ——————————————————————————————

This study found all viable collisions between pairs of fundamental particles A, B, C, D, E, F, Fbar, G, H. These do not include repeated or compound particles such as AA, ... FF, GG, G'G, HH, H'H, GH, HG, G'H, H'G, etc.

Some collisions were found to have varying resulting particles, due to their reactants interacting at different distances away from each other. Any collision between two fundamental particles in which the spacing between the particles doesn't affect the result, is denoted by "A_B" for any particles A and B. The "_" in between particles' names implies that the spacing in between the particles does not affect the outcome of the collision. However, in the collisions that have integers between the particles' names, the integer represents the number of "spaces" between the two particles in that specific collision. A "space" is a full spatial period of the background, and consists of 6 consecutive cells. The integer must first be evaluated in a modular function that is specific to the collision. Below is a table of all viable collisions between fundamental particles.

| Particle A | Particle B | Particle C | Particle D | Particle E |
|---|---|---|---|---|
| A_C | B_C | C_D | D_F | E0F |
| A_D | B_D | C_E | D0Fbar | E1F |
| A_E | B_E | C_F | D1Fbar | E2F |
| A_F | B_F | C_Fbar | D_G | E0Fbar |
| A0Fbar | B0Fbar | C_G | D_H | E1Fbar |
| A1Fbar | B1Fbar | C_H | | E2Fbar |
| A_G | B_G | | | E0G |
| A_H | B_H | | | E1G |
| | | | | E2G |
| | | | | E_H |

Figure 20: Table of all possible collisions between fundamentals.

*Discussion: Block Emulation* ───────────────────────────────────────────

Using Rule 20645 in the 3/2 rule space, we can identify several rules that Rule 73 can emulate through block emulation, with blocks ranging in size from 0 cell to 16 cells [16]. We must search the emulated rules for signs of universality, so let us identify the rule numbers for ECA Rule 110 and Rule 193 in the 3/2 rule space. Using the procedure in *Methodology: Block Emulation*, we find that Rule 110 is equivalent to Rule 23290, and Rule 193 is 61445.

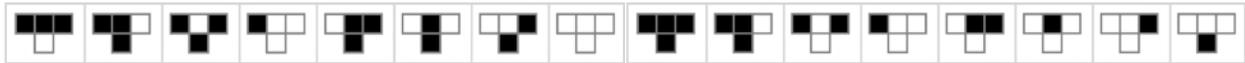

Figure 21: Rule 110 (left) and 193 (right) are equivalent and universal. They are Rule $23290_{3/2}$ and $61445_{3/2}$ respectively.

Here we can see all of the rules that Rule 20645 emulates up to a block size of 16 cells:

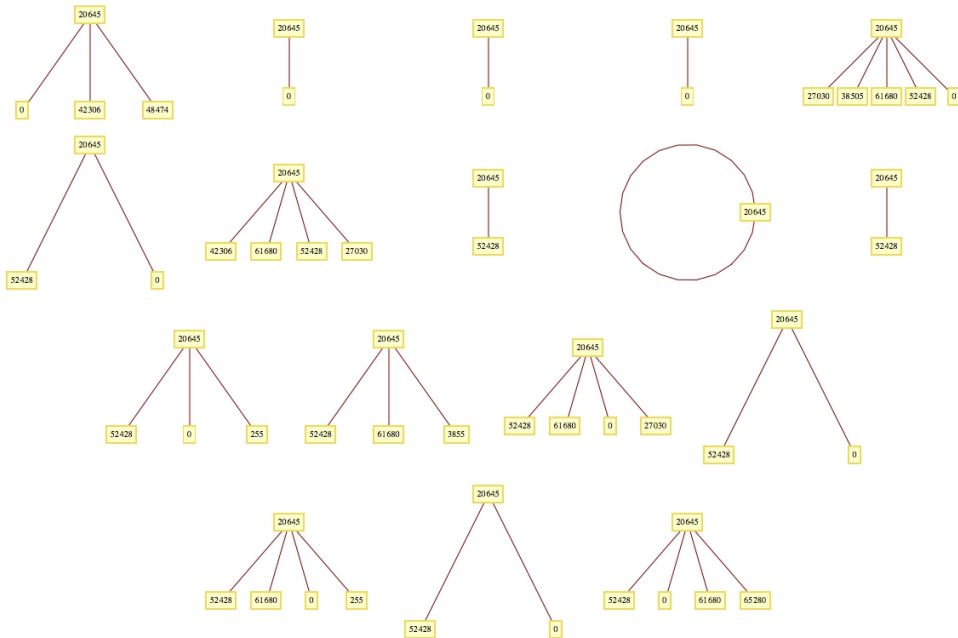

Figure 22: A diagram for all of the rules in the 3/2 rule space that Rule 20645 emulates up to a block size of 16 cells. Each tree of rules represents a different block size, ranging from 0 to 4 cells on the first row, 5 to 9 cells on the second row, 10 to 13 cells on the third row, and 14 to 16 on the last row [16].

This form of block emulation was continued until the block size was 30 cells, but none of the emulated rules were 23290 or 61445, so no desired result was generated. Additionally, as the emulation approached a block size of 40 cells, the estimated runtime jumped incredibly high, and the block emulation became limited by the amount of physical memory available. This prevented me from getting consistent results with any block larger than 30 cells, and greatly handicapped my ability to prove the universality of Rule 73 through this method.

*Discussion: Neighbor-Dependent Substitution System* ─────────────────

Despite the constraints of the previous method, we can still prove universality in a variety of other ways. For example, let us consider the neighbor-dependent substitution system [2]. A particle detector that was created for Rule 73 in LDCA-1-2 returns a string of particle names and background that resembles "A----B--------C-GG-----H-G" at any evolutionary step. Following suit, we can construct a neighbor-dependent substitution system that takes a similar string as an input, and substitutes pairs of colliding particles with their outputs. The program should continue to collide particles until there aren't any legitimate pairs of particles left, and then return a list of past states at every collision.

In the neighbor-dependent substitution system, the spacing between any consecutive particles in a collision was replaced with an integer signifying the number of background rows and evaluated with modular functions, so that large numbers of background cells wouldn't cause the program to malfunction. For every spacing-dependent collision, the substitution system would classify the result of the collision using the integer between the particles, and replace the pair with a corresponding result sandwiched between two additional spacing values to account for background that was lost as the string was modulated. Then, after every substitution, any adjacent numbers of background cells in the string were be added so that they behaved as a singular spacing. The modular function that was applied to the spacing was either mod 2 or mod 3, depending on the pair of particles involved.

Unfortunately, this neighbor dependent substitution system was inconclusive. The set of rules that dictated the substitutions in the system had to stay restricted to usable collisions, since some collisions resulted in a particle generator that created an infinite number of products. However, the rules were unable to stay restricted as particles that had been excluded from the list eventually resulted from collisions, and were added to the rule set. New collisions had to be added to the substitution system, and ultimately, the neighbor-dependent substitution system failed to simplify any string of particles.

*Discussion: Collision System* ─────────────────────────────────────────────

Finally, let us turn to collision G'G_B'B which returns particles B'B and G, and collision G_B'B which returns particle G'G. Using these collisions, it is not difficult to construct a system that consists of two different substitution rules {{AB → $B_1C$}, {CB → A}}, and plot the behavior of said system in which alternating rows of G'G and G are colliding with B'B.

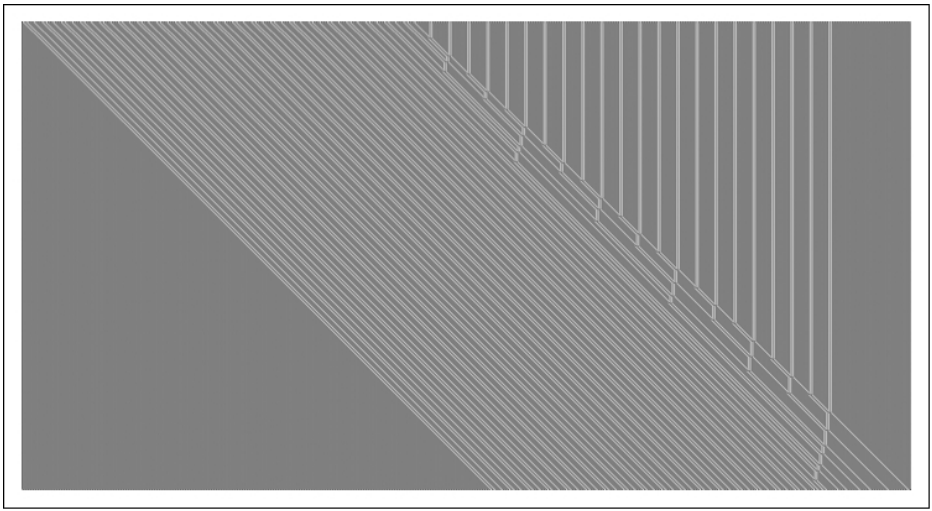

Figure 23: A chain of collisions between alternating rows of G'G and G, and B'B [4]. The system consists of two different substitution rules {{AB → $B_1C$}, {CB → A}}.

Additionally, in every collision, the G or G'G output is released with a 3-cell shift to the right, while the B'B output is shifted 6 cells to the left. This means that for every pair of G_B'B and G'G_B'B collisions, B has an overall shift of 6 cells to the right, and the resulting G particle is shifted 6 cells to the right.

As is observable, there are 3 points in total (on Figure 17) where all the particles that are colliding with B'B are Particle G'G (except for the G that annihilates the B'B at the end), and they are all converted at once into Particle G. At those points, the first B'B converts 3 G'G particles, the second converts 5, and the third converts 7. These "G'G conversion" collisions count consecutive odd numbers as the system progresses, in the pattern 3, 5, 7, … 2n+1. Additionally, counting the collisions between the G'G conversion points reveals interesting results, as we can see in this chronological list of collisions where each number represents how many collisions the B'B particle endures at that relative location ("*" is a G'G conversion point):

1 2 1 * 1 2 1 3 1 2 1 * 1 2 3 2 1 4 1 2 3 2 1 * 1 2 3 4 3 2 1 5 1 2 3 4 3 2 1 * …

However, we notice that this string of numbers can be thought of as an inorder traversal of a series of binary trees [17]. And, when counting the number of collisions in which B'B interacts with k number of particles, we find that the sums are of the form $2^{n-k+1}$. For example, the total number of collisions in which B'B interacts with 1 particle, in between G'G conversion points, follows the pattern 2, 4, 8, … $2^n$.

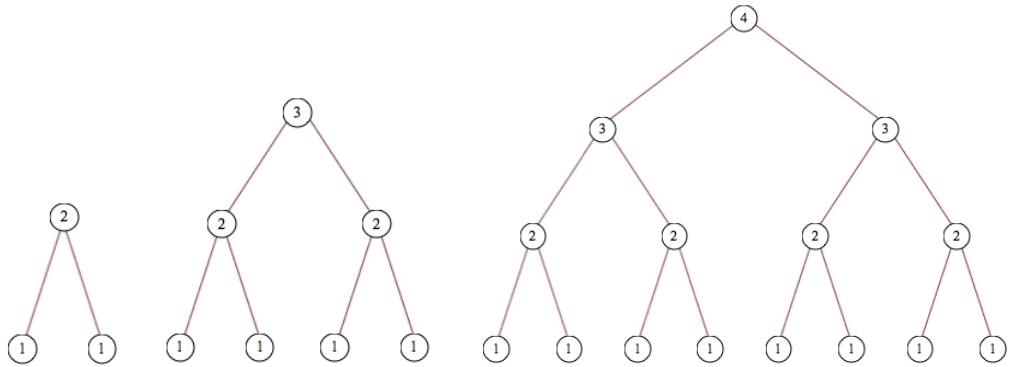

Figure 24: Complete binary that dictates the progression of collisions in the G'G and G, and B'B collision system [17].

We also find that when counting the *total* number of interactions of any type in between the G'G conversion points, one is left with a progression of Eulerian numbers, which follows the series, 4, 11, 26, 57, 120, 247, 502, … [19][20]. Interestingly though, the interactions between G'G conversion points don't define the standard Eulerian numbers of Series A008292 [18][21], but instead correspond to the values of the second column (k = 2) of the standard Euler Triangle (which define series A000295) [20]. Thus, through the G, G'G, and B'B collision system, sequence A000295 is inherently related to the mathematical properties of complete binary trees. This ability of Rule 73 in LDCA-1-2 to associate the behaviors of binary trees with Eulerian numbers (in series A000295 of [20]) can in turn provide valuable insight to unsolved problems such as those in [22], and lead to future mathematical exploration.

*Future Research* ─────────────────────────────────────────────

In future research, it is suggested that the connections between binary trees and Eulerian numbers be studied through the use of Rule 73, and not just mathematical expression. With a LDCA approach, one may be able to better understand, both technically and conceptually, the mathematical nature of both constructs, and gain an insight into problems similar to those presented by Baril and Pallo [22]. By utilizing the unique properties of Rule 73 in LDCA-1-2, it is possible that further mathematical exploration could prove more fruitful than previously expected.

In addition to the above, the compound particles of Rule 73 should be studied in more detail, so that a functioning neighbor-dependent substitution system might be generated. With more complex gliders, the behavior of collisions may be diverse enough that a defined set of rules can be used to evolve and simplify a string of particles, which would mean that a neighbor-dependent substitution system could become a viable option to emulate other universal systems.

And, the block emulation of Rule 73 as Rule 20645 in the 3/2 rule space will be more feasible to study in the future. While the scope of this study was limited by physical constraints, it is likely that future attempts at proving the universality of Rule 73 could make more progress with the block emulation of Rule 20645 with further code optimization and improved hardware, and succeed in emulating rules with block sizes much larger than 6 cells [16].

But besides making progress on methods already used in this study, any more research that is done on Rule 73 in LDCA-1-2 should have an emphasis on the emulation of a cyclic tag system. In a cyclic tag system, a standard tag system is applied to an initial condition in sequential order, and in a cyclic fashion that loops back to the front of the tape [14]. The discovery of the G, G'G, and B'B collision system is a large step forward in the process of emulating a cyclic tag system with Rule 73, since it contains types of collisions that were used in Cook's proof of universality for Rule 110 [14].

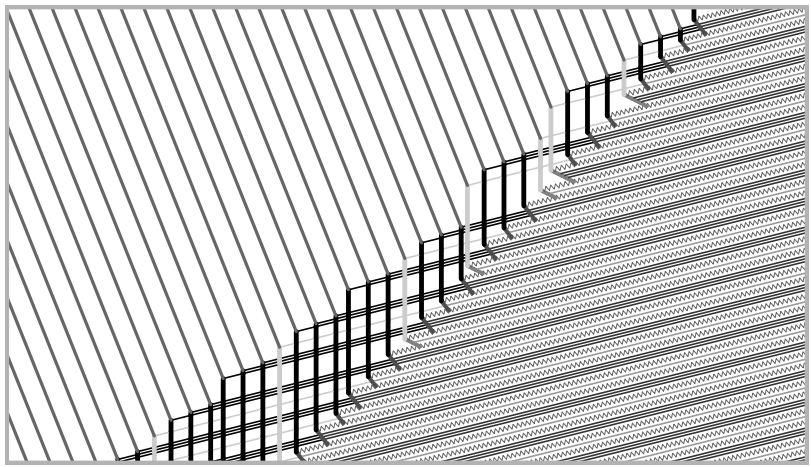

Figure 25: An example of a cellular automaton (Rule 110) emulating a universal cyclic tag system [14]. Some of the collisions represented to the left bear resemblance to the G'G, G, and B'B collisions that were found in this study.

Finally, it is advised that more research is conducted on the properties of LDCA. The effects of changing Rule 73's configuration from LDCA-1-1 (ECA) to LDCA-1-2 have been significant, and have turned Rule 73 into a potentially universal cellular automaton. In the future, additional research should be conducted on the effects of different LDCA configurations on universality. Hopefully, the exploration of LDCA will bring to light new possibilities and help further our current knowledge of evolving computation systems [23][7].

By exploring Rule 73 in LDCA-1-2, uncovering its applications to problems in other fields of research, and showing its candidacy for a new type of computational universality, I hope for the expansion and improvement of modern computation and mathematical modeling.

*References and Comments* ───────────────────────────────